\documentclass{ws-mplb}
\usepackage[super]{cite}
\usepackage{graphicx}
\usepackage{amsmath}

\begin{document}

\markboth{O. V. Pavlovsky, V. I. Dorozhinsky, S.D. Mostovoy}
{Artificial stochastic neural network on the base of double quantum wells.}

%%%%%%%%%%%%%%%%%%%%% Publisher's Area please ignore %%%%%%%%%%%%%%
\catchline{}{}{}{}{}
%%%%%%%%%%%%%%%%%%%%%%%%%%%%%%%%%%%%%%%%%%%%%%%%%%%%%%%%%%%%%%%%%%%

\title{ARTIFICIAL STOCHASTIC NEURAL NETWORK ON THE BASE OF DOUBLE QUANTUM WELLS.   
}

\author{\footnotesize  O. V. PAVLOVSKY }

\address{Faculty of Physics,
Moscow State University,\\
Moscow, 119991, Russia\\
Institute for Theoretical and Experimental Physics,\\
Moscow, 117218, Russia \\
pavlovsky@physics.msu.ru}

\author{V. I. DOROZHINSKY}

\address{Faculty of Physics,
Moscow State University,\\
Moscow, 119991, Russia\\
dorrozhin@gmail.com}

\author{S. D. MOSTOVOY}

\address{Faculty of Physics,
Moscow State University,\\
Moscow, 119991, Russia\\
Institute for Theoretical and Experimental Physics,\\
Moscow, 117218, Russia \\
smostovoy@mail.ru}

\maketitle

\begin{abstract}
We consider a model of an artificial neural network based on quantum-mechanical particles in  $W$ potential. These particles play  the role of neurons in our model. To simulate such a quantum-mechanical system the Monte-Carlo integration method is used. A form of the self-potential of a particle as well as two interaction potentials (exciting and inhibiting) are proposed. Examples of simplest logical elements (such as AND, OR and NOT) are shown. Further we show an implementation of the simplest convolutional network in framework of our model. 

\end{abstract}

\ccode{PACS Nos.: 03.65.-w., 07.05.Mh}

\section{Introduction}

The modern development in various fields of science and technology
strongly depends on progress in computer science that
may be associated with the implementation of new computing system technologies as well as new computational algorithms.
Quantum computing and artificial intelligence seem to be the two major tendencies in the present development of the computer
science.

In the course of time logical elements of computer central processors are becoming smaller and smaller. Nowadays technologies make it possible to produce the elements so tiny that quantum fluctuations become more and more involved. On the one hand, the
quantum nature of  such objects can lead to the crisis of the
classical computing technique, but on the another hand, their
quantum properties can be used to implement quantum computation algorithms.

As noted above, the quantum limit significantly limits the size of the basic elements of computer microprocessors. It turns out to be a significant limiting factor that influences the development of modern electronics.

Artificial intelligence is another main tendency in modern development of the computer science.  Artificial neural networks  are computing systems inspired by the biological neural networks and
are composed from calculation nodes (artificial
neurons) connected to each other. The connections between such nodes
transmit  the  signals from one node to another as neurons in a brain do.  The parameters of the connections depend on weights, the change of which modifies the transport of the signals in the neural
network. The learning procedure of the neural network is just
reduced to tuning these weights. A number of
computing problems (for example, image \cite{lecun} and speech \cite{speech1} recognition, machine translation \cite{google}) can be solved with help of artificial neural networks.

As was mentioned above, the quantum limit significantly limits the size of basic elements of computer microprocessors. Now, one can ask how small the artificial neural networks could be. Additional issues concern the working regime near the quantum limit and a valid implementation of neural networks with help of quantum physics. In this paper we will show that quasi-classical  fluctuations of some quantum systems can assist in the artificial stochastic neural networks consideration \cite{kappen}.

The main idea of our work is that semi-classical fluctuations of quantum systems can be used to implement a stochastic neuron. These semi-classical fluctuations arise, for example, when a quantum particle tunnels from one quantum well to another. In our paper a quantum mechanical model of the stochastic neural network such as double quantum wells will play the role of stochastic neurons.

The basic principles of stochastic neural networks are known long ago \cite{old1, old2, old3, old4, old5, old6}. In contrast to an ordinary (classical) neural network with neurons whose neuron function is deterministic, the stochastic neuron network consists of neurons with a stochastic neuron function. The probability of the spike generation in response to some input signals is considered alone. So the whole stochastic neural network works as a statistical machine. Thereby, the stochastic neural network can reveal a typical statistical phenomenon like the phase transitions and the strongly-correlated regimes \cite{strong}. In biological neural networks such strongly-correlated regimes can be associated with the cognitive processes in the brain. These regimes can play an essential role in learning of the neural networks. All these facts make the study of stochastic neural networks an extremely interesting and actual task.

Historically, stochastic neural networks have been inspired by thought processes examination in a brain \cite{cowan1, cowan2}. However, nowadays the ideas of stochastic neural networks were transferred to the computer science too. It has been shown that stochastic neural networks can be applied to solution of many important computational problems \cite{new1, new2, new3}.

An implementation of the stochastic neural networks is actually related to the numerical simulation of these networks on ordinary computers. Therefore it is a very important question how one can practically produce stochastic neural networks? In our work we suggest that such stochastic neural networks can be created on the basis of quasi-classical processes in double quantum wells. It is not a novel idea to use double quantum wells in order to build computers (classical or quantum). It has been proposed to design quantum gates \cite{ddot} in terms of the wells. It is possible to use the quantum dots \cite{dot} for the quantum neural networks \cite{kek}. In our work double quantum wells are employed to create stochastic neurons.

An additional motivation for our work was the recent creation of a neuromorphic network based on nanowires \cite{wire}. This brain-inspired network implements so-called edge-of-chaos learning. This type of learning is an essential concept for building neural networks capable of self-adaptation and possibly takes place in the human brain \cite{edge-of-chaos_brain}. Also it is important to note that a self-adaptive neural network was recently created \cite{bp_bm} by using cobalt atoms on a semiconductor black phosphorus substrate \cite{bp_co}. We aimed at creating a theoretical model of a neural network that could be simply investigated numerically, so that the phenomena of the edge-of-chaos learning and the self-adaptation could be studied.

The paper is organized  as follows. We start with discussion of a quantum-mechanical model of a single stochastic neuron.
%Any neuron must generate spikes of activity spontaneously, or under the influence of external influences. For numerical simulation of all quantum-mechanical models the Path Integral Monte Carlo method was used. 
The stochastic neural networks consist of numerous stochastic neurons which are connected to each other. In the second part of our work we introduce a possible way of such connection by means of a specific interaction between neurons. Next, we consider some applications of this stochastic neural network to implement the logical elements and the convolutional network for vertical line detection.

\section{Stochastic neurons as double quantum wells}

\subsection{Single stochastic neuron}

The main aim of a neuron (classical and stochastic) should be
to produce spikes or bursts of activity. The integrate-and-fire neuron is one of the simplest models. This simplification of the neuron's function is based on
the integration of external activity of other neurons to
produce a spike if this integrated external activity exceeds a certain threshold value. In this scenario the neuron excitation is transmitted to other neurons in the network. 

In our approach the double quantum wells play the role of stochastic neurons. The quantum objects possess an incorporated stochasticity and so one can only discuss a probability of the particle excitation in the quantum well. However the quantum nature of our artificial neurons will not preclude basic functions of the neural networks to be revealed because we use a spatial feature of a double quantum well systems. The tunneling processes between the wells in such systems can be connected to the quasi-classical fluctuations and so one can consider them as a stochastic analog of spike activity in classical neurons. 
 
Let us consider the simplest one-dimensional model of a double quantum well with the Hamiltonian
 
 \begin{equation}
\hat{H}_{i} = \frac{1}{2} \hat{p}_{i}^{2} + V_{0} \left( \hat{\varphi}_{i} \right),
\end{equation}
where $\hat{p}$ $\hat{\varphi}$ are quantum operators of momentum and coordinate. The potential energy $V_{0}$ is 
\begin{equation}
V_0(\hat{\varphi}_{i}) = \frac{\Lambda}{4}\left( \hat{\varphi}_{i}^2 - 1 \right)^2.
\end{equation}

This is a well-known and thoroughly studied $\varphi^4$ model in one dimension. Two types of classical solutions with a finite value of classical action exist: classical vacuums $\varphi_0 = \pm 1$ and so-called instantons $\varphi_I = \pm \tanh (\sqrt{\lambda/2}(t-t_0)) $ \cite{polyakov}.  The qualitative picture of behavior of the quantum system can be thought of as small quantum fluctuations around the classical vacuums and a spontaneous tunneling which corresponds to instantons.

\begin{figure}
\includegraphics[width=0.99\linewidth]{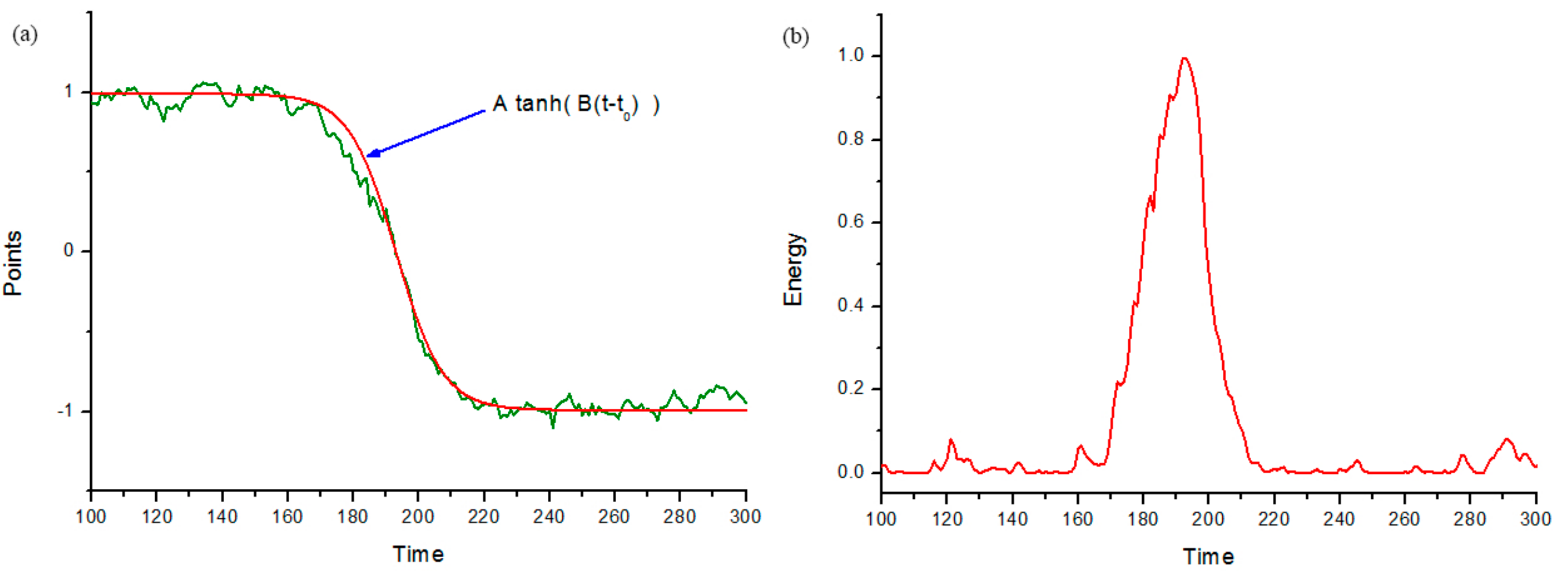}
\caption{(a) Euclidean trajectory $\varphi(t)$ with an instanton and (b) corresponding density of the potential energy $V_0 (\varphi(t))$.} \label{fig:instanton}
\end{figure}

In Fig.~\ref{fig:instanton} one can see the Euclidean trajectory $\varphi(t)$ with an instanton and corresponding density of the potential energy $V_0 (\varphi(t))$. This trajectory was generated by means of Path Integral Monte-Carlo method which is discussed in the next Section. In Fig.~\ref{fig:instanton}(a) the classical instanton solution (red line) for the illustration is shown.
%how the tunneling processes can be  found on the graph.
It is very important to point out that the density of the potential energy $V_0 (\varphi(t))$ has a peak just at the moment of tunneling near the instanton center. We will use this effect in the design of the stochastic neurons. Such instanton peaks can play a role of a spike which is an analog of biological spikes in the brain neurons.

In our approach we will use the density of the potential energy $V_0 (\varphi(t))$ as an estimation of the network node's activity.  If the node is not active, then the trajectory of the node fluctuates around the classical vacuum. A high node activity means a large number of instantons on this node. From a physical point of view, this leads to a high average energy density in this mode which can be detected by an external observer.

In this Section we consider only one node of our artificial neural network. Instantons (or stochastic spikes) on each node (or neuron) arise spontaneously and the goal is to link the network nodes so that the instanton (spike) on one node generates instanton (spike) on the next network node. If we organize the connection between neurons in a proper way, then the activity will be transferred from one part of the network to another and our neural network  will start to work.

In the next Section the Path Integral Monte-Carlo method is briefly discussed and its application to study of the quantum-mechanical models of the stochastic neural network is developed.

\subsection{Path Integral Monte-Carlo method}

In the previous Section a single stochastic neuron was discussed. This system can be studded analytically by means of quasi-classical expansion method. But the ultimate goal is to study a system of many such quantum nodes with complicated connections to each other. There is no way to fulfil this task analytically and the numerical methods should be involved. An extremely effective method for studying large quantum systems is the Path Integral Monte-Carlo method \cite{ceperley}. 

Let us consider a model of a quantum mechanical system organized in a neural network in terms of the Path Integral formalism. 
A neural network consists of nodes (neurons) which are one dimensional quantum mechanical systems. These nodes are connected to each other by an interaction potential which plays the role of axons
\begin{equation}
V_{int} = V_{int} \left( \hat{\varphi}_{i}, \hat{\varphi}_{j} \right).
\end{equation}
Thus, the total Hamiltonian of the system reads
\begin{equation}
\hat{H} = \sum_{i}\left( \frac{1}{2} \hat{p}_{i}^{2} + V_{0}( \hat{\varphi}_{i} )\right)
+ \sum_{i > j} V_{int}( \hat{\varphi}_{i}, \hat{\varphi}_{j})\\
\end{equation}
where index $i$ enumerates nodes in the network. 

In the general case we need to deal with sufficiently
complex quantum-mechanical systems, so a well-known Path Integral approach is to be used to describe their properties. In Euclidean time the statistical sum of the system has the form
\begin{equation}
Z = \int \prod_{i} \mathcal{D} \varphi_{i}\left( \tau \right) \exp(-S(\varphi_{i}(\tau))), \,\,\, \varphi_{i}(0) = \varphi_{i}(T),
\end{equation}
where $\varphi_{i}(\tau)$ is the Euclidean path of $i$-th node, $\tau \in \left[ 0, T \right] $
is Euclidean time, and $S(\varphi_{i})$ is the classical action:
\begin{equation}
S = \int_{0}^{T}d\tau \left[ \sum_{i}\left(  \frac{1}{2} \dot{\varphi}_{i}^{2}
+ V_{0}( \varphi_{i} ) \right) %\right.\\
+ \sum_{i > j} V_{int}( \varphi_{i}, \varphi_{j}) \right].
\end{equation}\par
The observables in such formalism are calculated by
\begin{equation}
\mathcal{O}(\varphi_{1},...,\varphi_{i}) =
\frac{1}{Z} \int \prod_{i} \mathcal{D}\varphi_{i}(\tau)\mathcal{O}(\varphi_{1},...,\varphi_{i})\exp(-S(\varphi_{i})).
\end{equation}

The operation of the network is based on the propagation of
activity from the input nodes (sensors) to the output ones. As
already stated, the network can consist of numerous nodes. To study
such a complicated quantum system, it is natural to apply the Monte
Carlo method\cite{ceperley}. The major idea is to use a
Markov process to generate paths of particles $\varphi_{i}$ with
a statistical weight proportional to $\exp(-S(\varphi_{i}))$.\par
In our work we use the multilevel algorithm of Metropolis. An 
introduction of several levels of the algorithm is motivated by the need to suppress autocorrelations and makes it possible to improve the computational efficiency.

In our calculations the parameters of the Metropolis algorithm were chosen as follows:
time (inverse temperature) $T = 0.7$, number of time grid nodes
$N_t = 512$. To suppress autocorrelations we use the
thermalization of $ 2\cdot10^6$ iterations long.

%Finally, note
%that for the initialization of the path we use the saw path of 0's
%and 1's: $\varphi_{init}=i \bmod 2$ where $i \in [0, 512] \cap
%\mathbb{Z}$ is the number of corresponding time grid node.

\section{Excitatory and inhibitory connections of neurons and logical elements}

\subsection{Single neuron in sleeping state}

\begin{figure}
\centering
\includegraphics[width=\linewidth]{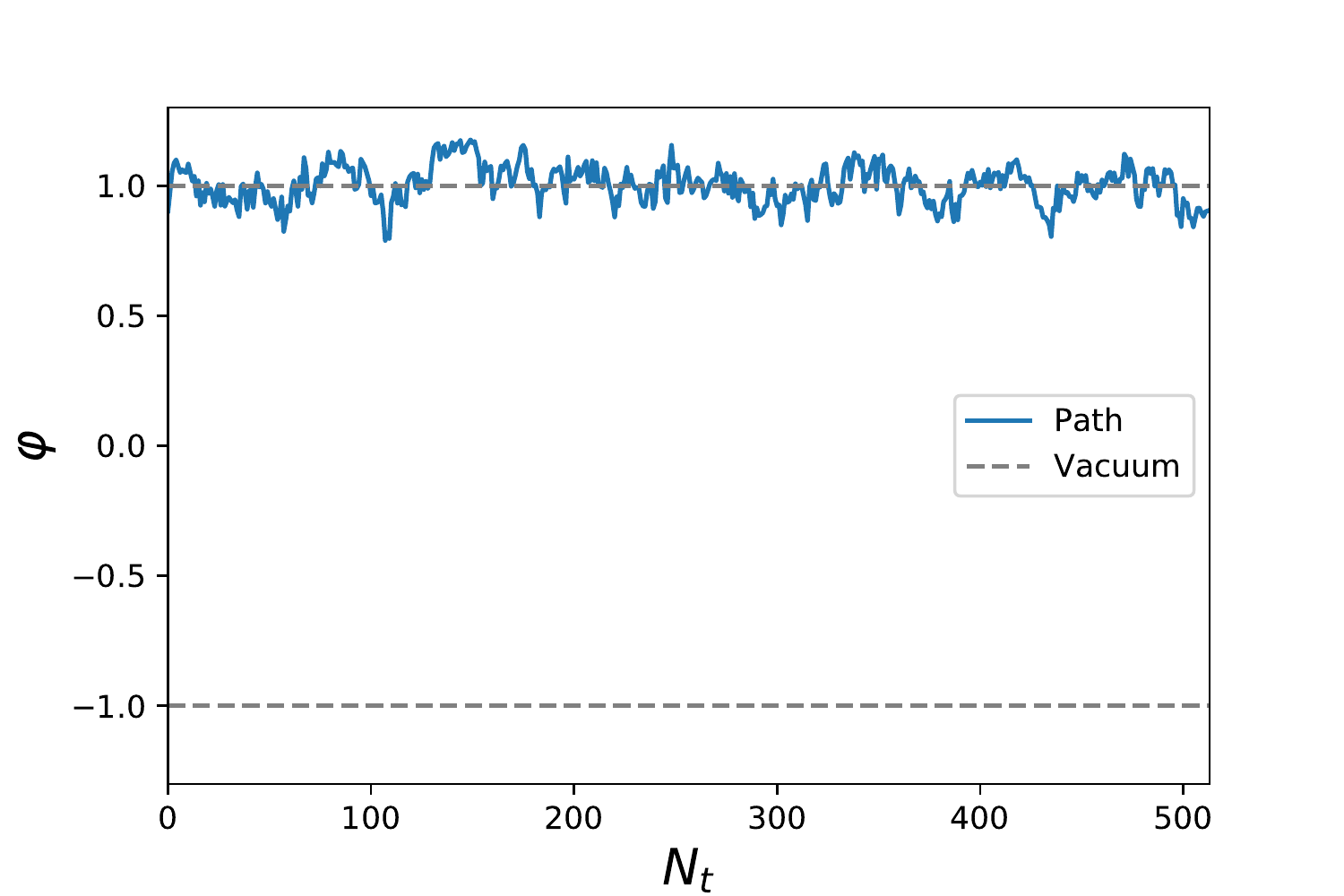}
\caption{Free neuron sleeping state. The neuron will remain in the vicinity of one of the vacuums (here $\varphi = +1$) until an external influence occurs.}
\label{fig:singleNeuronPath}
\end{figure}

We start with a single neuron. Let us consider an artificial neuron as a quantum particle in a $W$ potential. The sleeping (or rest) state of the neuron is represented by fluctuations about the classical vacuums $\varphi = \pm 1$ and can be seen in Figure~\ref{fig:singleNeuronPath}. A spike of neuron activity is connected with the tunneling process of the particle from one vacuum to another. 

The Lagrangian of such a system can be written as:
\begin{equation}
\mathcal{L}_{0} = \frac{1}{2}\dot{\varphi}^{2} + \frac{\Lambda}{4} \left(\varphi^2 - 1 \right)^{2}.
\end{equation}

For the neural network to operate correctly a single neuron without connections with any other neurons of the network must be at a sleeping state (Figure~\ref{fig:singleNeuronPath}). A neuron must generate the spike only as the result of the external influences. Therefore, it is necessary to choose the values of the action parameters for the single neuron (or potential $V_0 (\varphi(t))$) so that spontaneous tunneling processes will be suppressed. The physically specific form of the potential depends on the form of the quantum well. The choice of the parameters should, however, allow a relatively small external influence to cause a response spike though.

In quasi-classical consideration, the probability of the tunneling processes depends on the value of the action on the instanton solution
\begin{equation}
S_{cl} = \frac{2\sqrt{2\Lambda}}{3}. \label{kink_act}
\end{equation}
Thus the value of parameter $\Lambda$ will be chosen so that the spontaneous tunneling are suppressed only a little -- the neuron must be ready to respond to an external stimulus with a spike. By means of numerical study we have found the optimal value parameter $\Lambda$ to be about  $5000$.
 
% A typical thermalized path is presented in the
% Fig.~\ref{fig:singleNeuronPath}.

\subsection{Two neurons connected to each other}

We now turn to the case of two interacting neurons. We want a
spike (or tunneling process) in one of them to cause a spike in the other, but spikes in the second one should not manifest a back affect upon the first one. This means that the interaction part of the Lagrangian of such neurons must be asymmetric (we call such type of connection an \textit{excitatory} connection). Let us consider the following form of the interaction part of the Lagrangian
\begin{equation}
\mathcal{L}_{int} =
\varepsilon_{exc}\varphi_{1}^{2}\left(\varphi_{2}^2 - 1\right)^2,
\end{equation}
where $\varepsilon_{exc}$ is the connection strength. If
$\varphi_{1}$ is in the vacuum, then there is no impact on
$\varphi_2$. However, if $\varphi_{1}$ experiences a spike, then
$\varphi_{2}$ also tends to have a spike. Thus a pulse is
spreading from one node to another. We will test it by means of a Monte-Carlo simulations.

%\begin{figure}
%\centering
%\includegraphics[width=\linewidth]{excition.pdf}
%\caption{Interaction potential for transmission of spikes from neuron %$\varphi_2$ to $\varphi_1$.}
%\label{fig:simpleExcitatory}
%\end{figure}

\begin{figure}
\centering
\includegraphics[width=\linewidth]{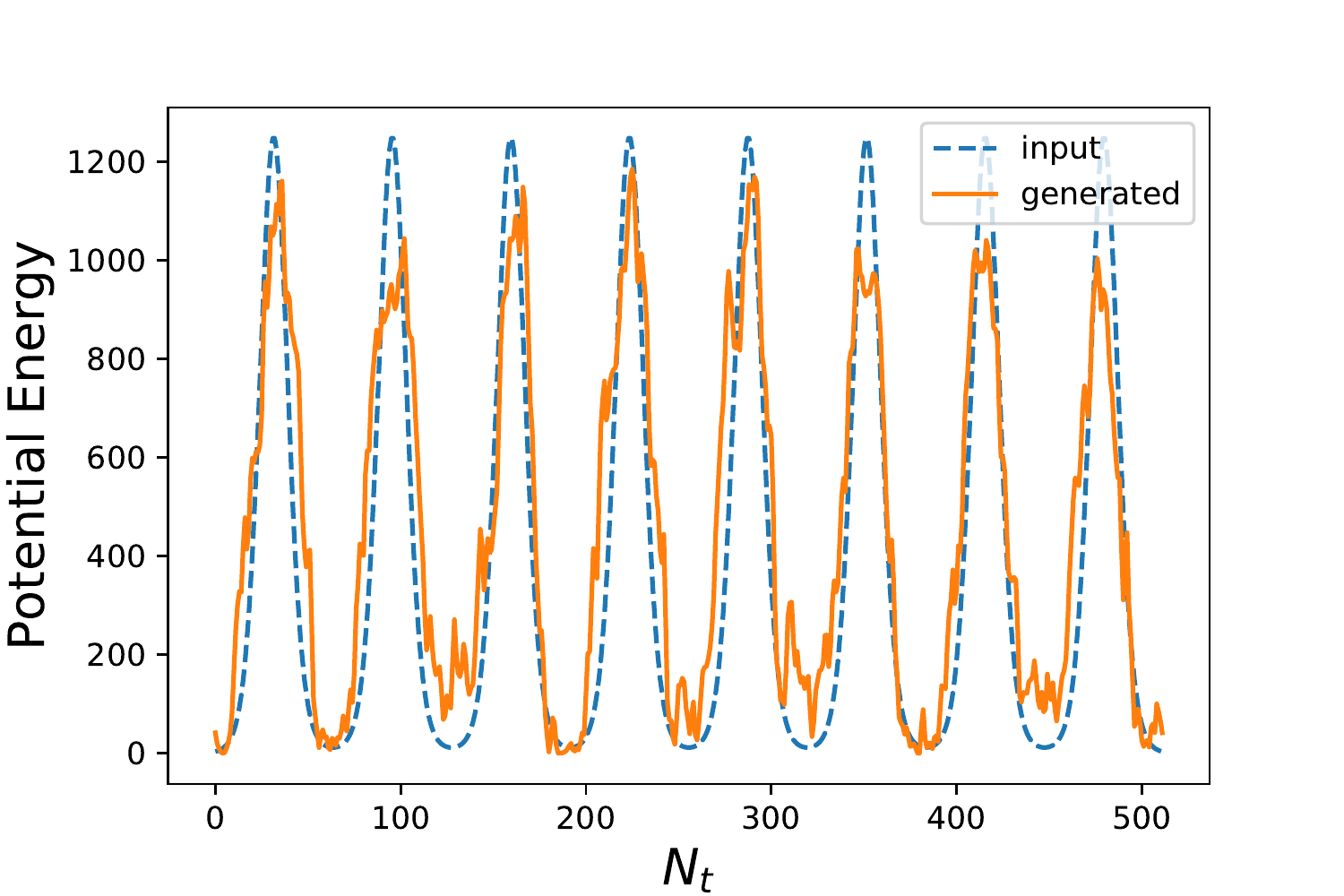} % second figure itself
\caption{Potential energy of an affected neuron (solid) tend to resemble the one of an input neuron (dashed).}
\label{fig:twoNeuronsConnected}
\end{figure}
%The plot of the potential of such interaction is shown in
%Fig.~\ref{fig:simpleExcitatory}. Axis $\varphi_2$ is related to
%the excitatory neuron and $\varphi_1$ corresponds to affected one.
%\par

The work of a neural network relies on its response to external information received from input neurons.
To simulate it, we use the input
neurons with some fixed activity. Each input neuron can be either passive (make no affect at all, may be discarded) or active. Active input neurons have a fixed path (unlike simulated neurons which path evolves during simulation) which consists of classical kink solutions. The potential energy of such an input neuron is depicted in Fig.~\ref{fig:twoNeuronsConnected} (dashed line). Each peak of the potential energy corresponds to a kink. Solid line represents the potential energy of a single neuron affected by an input neuron.\par

We introduce activity of any simulated neuron as a ratio of integral
potential energy of that neuron to the input (e.g. activity of
neuron which never leave vacuum will be $0$ and activity of a neuron
which path replicates path of input neuron will be $1$). In order to
investigate different schemes we will inspect plots of activity at
some neuron as a function of connection strengths
$\varepsilon_{exc}$.\par
\begin{figure}
\includegraphics[width=\linewidth]{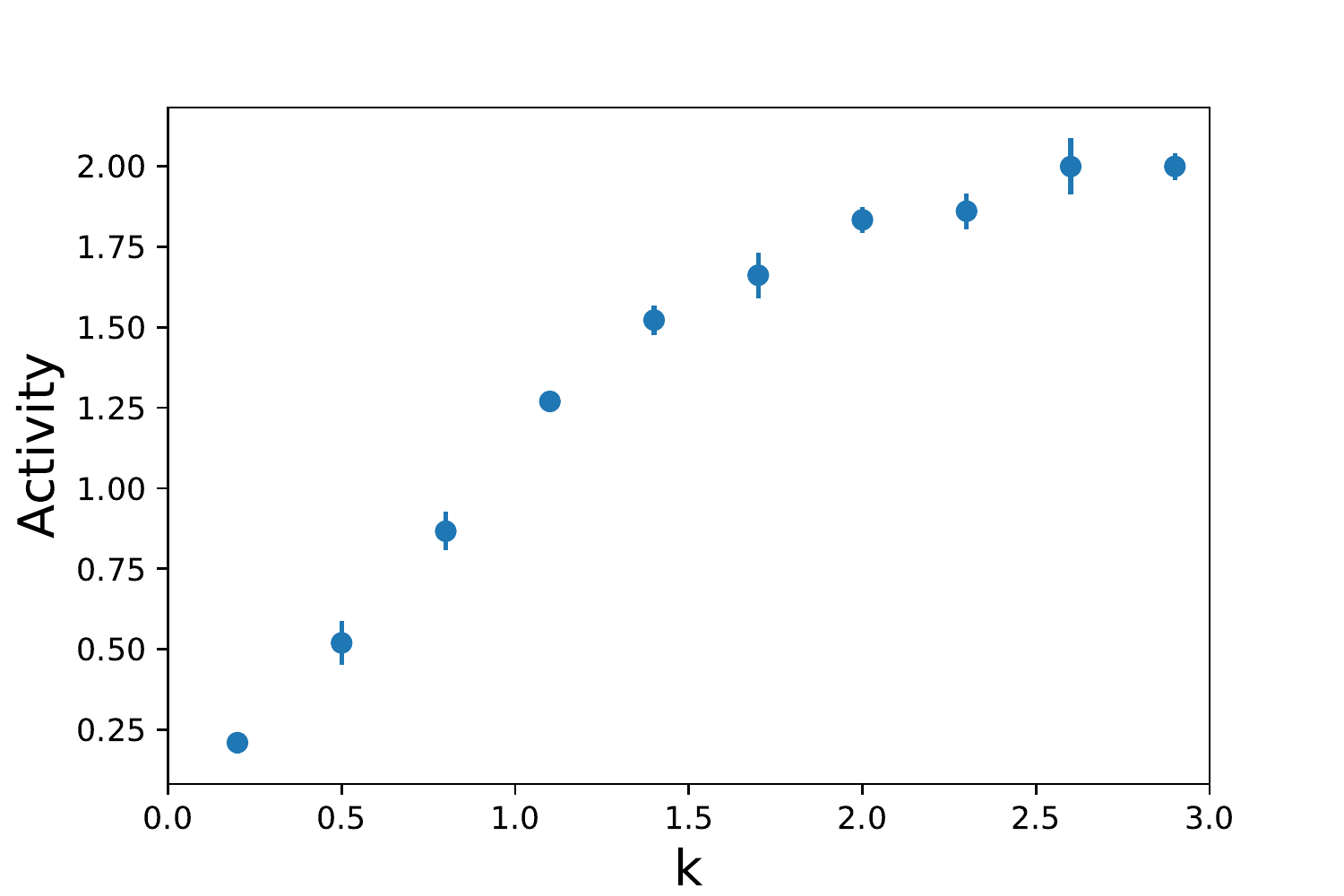}
\caption{
The activity of output neuron grows as a function of
connection strength.}
\label{fig:SIMPLE}
\end{figure}
In order to study different configurations of neurons we introduce
a modulating factor $k$. Once we choose appropriate parameter $\hat
\varepsilon$ for every connection, we multiply all of them by
this factor to obtain new connection strengths $\varepsilon = k
\cdot \hat \varepsilon$ and than plot the activity of neuron under consideration as a function of the parameter $k$.\par It was found
that $\varepsilon_{exc}$ can take it's values in the range from
$3000$ to $8000$ (Fig.~\ref{fig:SIMPLE}). In the case of too small
$\varepsilon_{exc}$, neurons almost do not interact and, if
$\varepsilon_{exc}$ is too large, their own potentials become
insignificant compared to the interaction, which leads to
an undesirable delay of the neuron in the state of $\varphi = 0$.
Fig.~\ref{fig:twoNeuronsConnected} presents a plot for $\varepsilon_{exc} = 6000$.\par

In the simulation presented in Fig.~\ref{fig:twoNeuronsConnected}
the activity of output neuron appeared to be 0.92.\par
\begin{figure}
\includegraphics[width=0.8\linewidth]{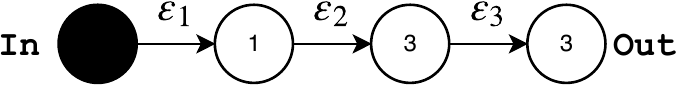}
\caption{The input neuron excites three simulated neurons placed in a
row.}
\label{fig:LineOf3}
\end{figure}

\begin{figure}
\includegraphics[width=\linewidth]{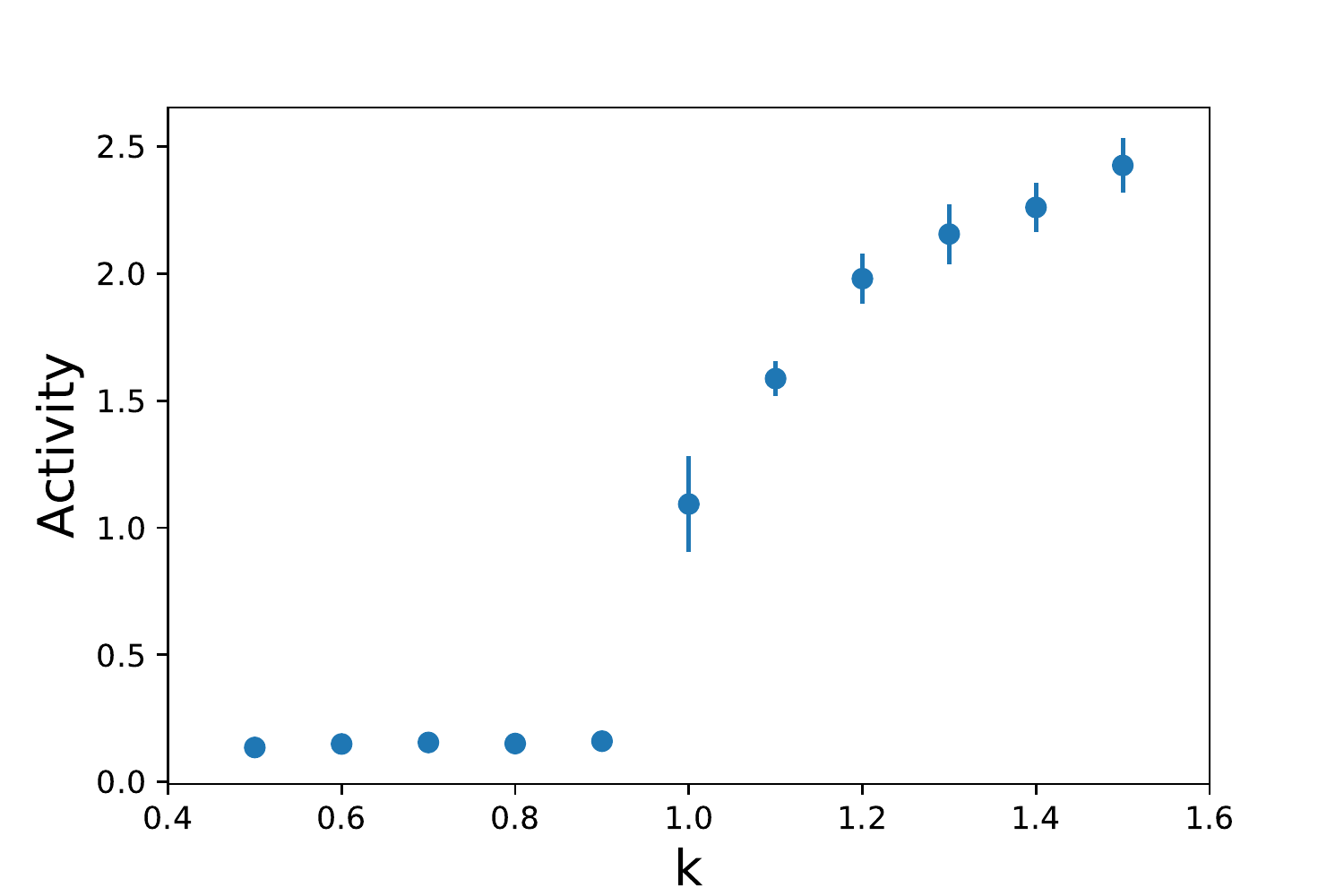}
\caption{The activity of output neuron (neuron 3) as a function of a connection strength. $\varepsilon_1 = k \cdot 1.5 \cdot 10^4,
\varepsilon_2 = k \cdot 1.0 \cdot 10^4, \varepsilon_3 = k \cdot
0.5 \cdot 10^4$. (see Fig.~\ref{fig:LineOf3})}
\label{fig:LINE_3}
\end{figure}

\begin{figure}
\includegraphics[width=\linewidth]{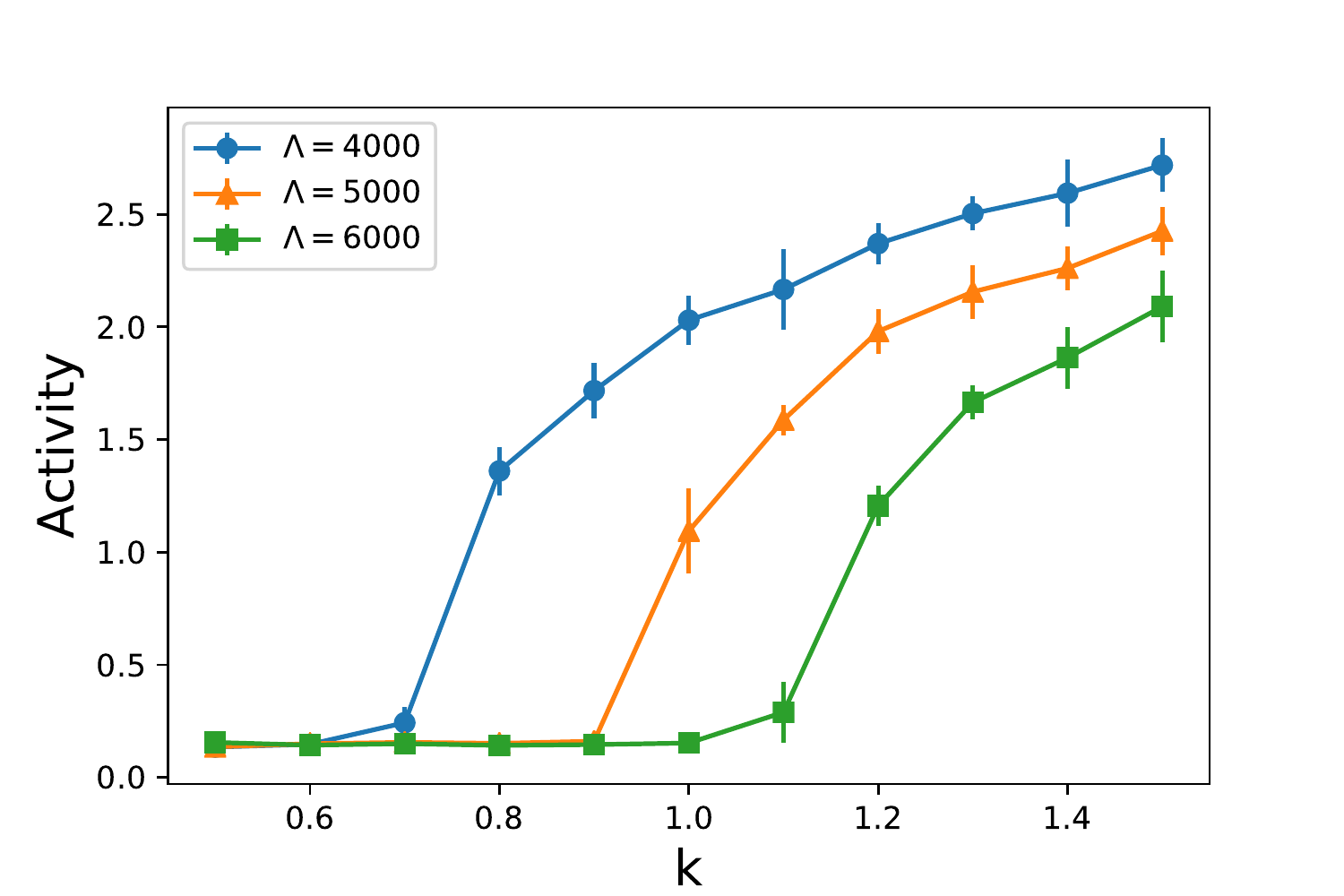}
\caption{The activity of output neuron as a function of $k$ for different $\Lambda$.
(see Fig.~\ref{fig:LineOf3})}
\label{fig:compare_lambda}
\end{figure}
It is possible to transmit an impulse through a line of several
simulated neurons. Consider the line of three simulated neurons
(Fig.~\ref{fig:LineOf3},~\ref{fig:LINE_3}). In this case we choose
$\varepsilon_1 = k \cdot 1.5 \cdot 10^4, \varepsilon_2 = k \cdot
1.0 \cdot 10^4, \varepsilon_3 = k \cdot 0.5 \cdot 10^4$. As one can
see from Fig.~\ref{fig:LINE_3},  for small values of the
connection strength $\varepsilon$ the spikes do not pass through
the chain of neurons, but if $\varepsilon$ reaches some critical
value, the chain becomes transparent for spikes. This effect
allows us to control the transparency of the neural network by a
slight change in the connection strength $\varepsilon$. Thus, we can realize complex
logical connections within our neural network by
controlling the connection strength.\par

The dependence of out neuron activity on the parameter $\Lambda$
is shown in Fig.~\ref{fig:compare_lambda}. It can be seen that the critical value of the connection strength $\varepsilon$ depends on $\Lambda$. Obviously, the obtained relation between the critical value and $\varepsilon$ is associated with an increase of kink's action (\ref{kink_act}).\par

\subsection{Logical elements}

\begin{figure}
\includegraphics[width=0.9\linewidth]{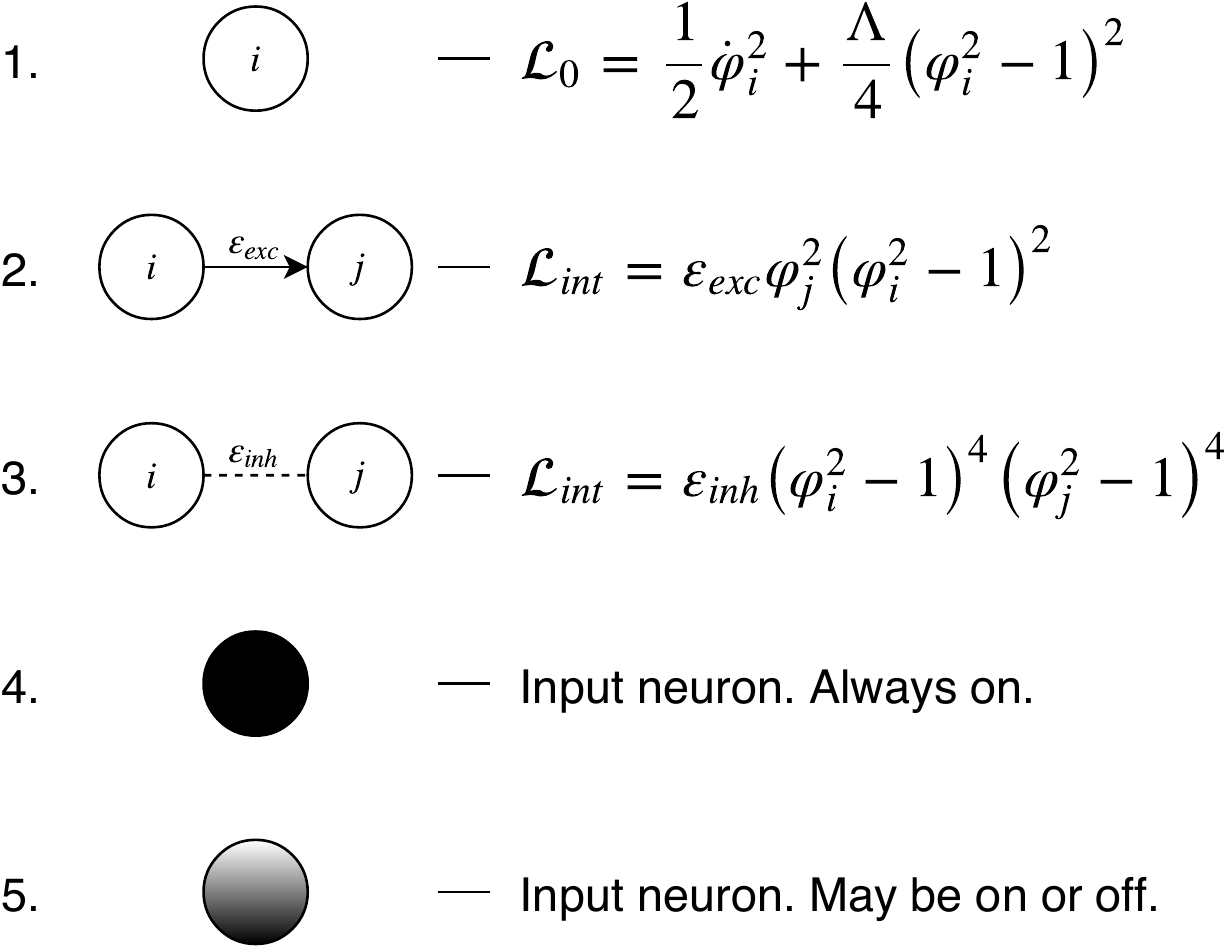}
\caption{Schematic diagram elements used to simplify the network representations. 1.
A contribution to the Lagrangian from neuron with index $i$. 2.
A contribution of an excitatory connection from neuron $i$ to $j$. 3.
A contribution of an inhibiting connection between neuron $i$ and $j$.
4. The input neuron that is always active. Its path does not change during a computer
simulation. 5. Input neuron that can be either in active or
passive mode. Depending on the state the
network should behave in different way. }
\label{fig:schemeSigns}
\end{figure}

We now turn to the construction of logical elements. In this
Subsection we construct from neurons introduced above the simplest
logical elements, such as AND, NOT, and OR. In order to
simplify presentation, we will use the schematic notation for
network elements (Fig.~\ref{fig:schemeSigns}).\par

\subsubsection{Logical AND}

Logical AND appears to be the simplest element to
construct. A neuron connected via this element with a set of
other neurons should experience a spike when all of the neurons it is
connected with experience a spike.\par To implement such a
behavior, it is necessary to employ an excitation
potential to connect the neurons whose signals are needed to be logically multiplied together. One should choose $\varepsilon_{exc}$
small enough to activate the output neuron only when all of
its inputs are activated, while it remains passive if at least one of the input neuron is passive.\par
\begin{figure}
\includegraphics[width=0.7\linewidth]{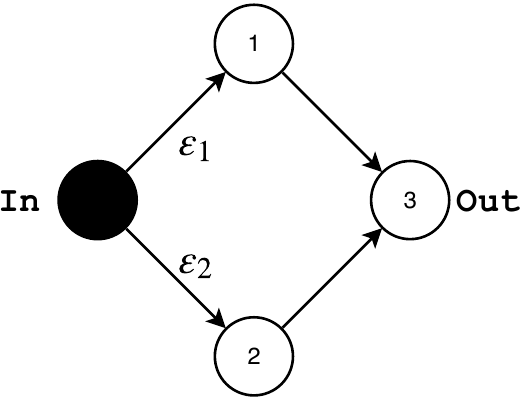}
\caption{A scheme implementing logical AND. Neurons 1 and 2 consume
input and neuron 3 receives the output information.}
\label{fig:andDemo}
\end{figure}
We demonstrate the operation of such a construction in the following
example. In Fig.~\ref{fig:andDemo} a scheme of a pictorial network is presented. A solid circle indicates an active input neuron. Circles with numbers depict simulated neurons. The arrows signed with values of $\varepsilon_{exc}$ show the connections.\par
\begin{figure}
\includegraphics[width=\linewidth]{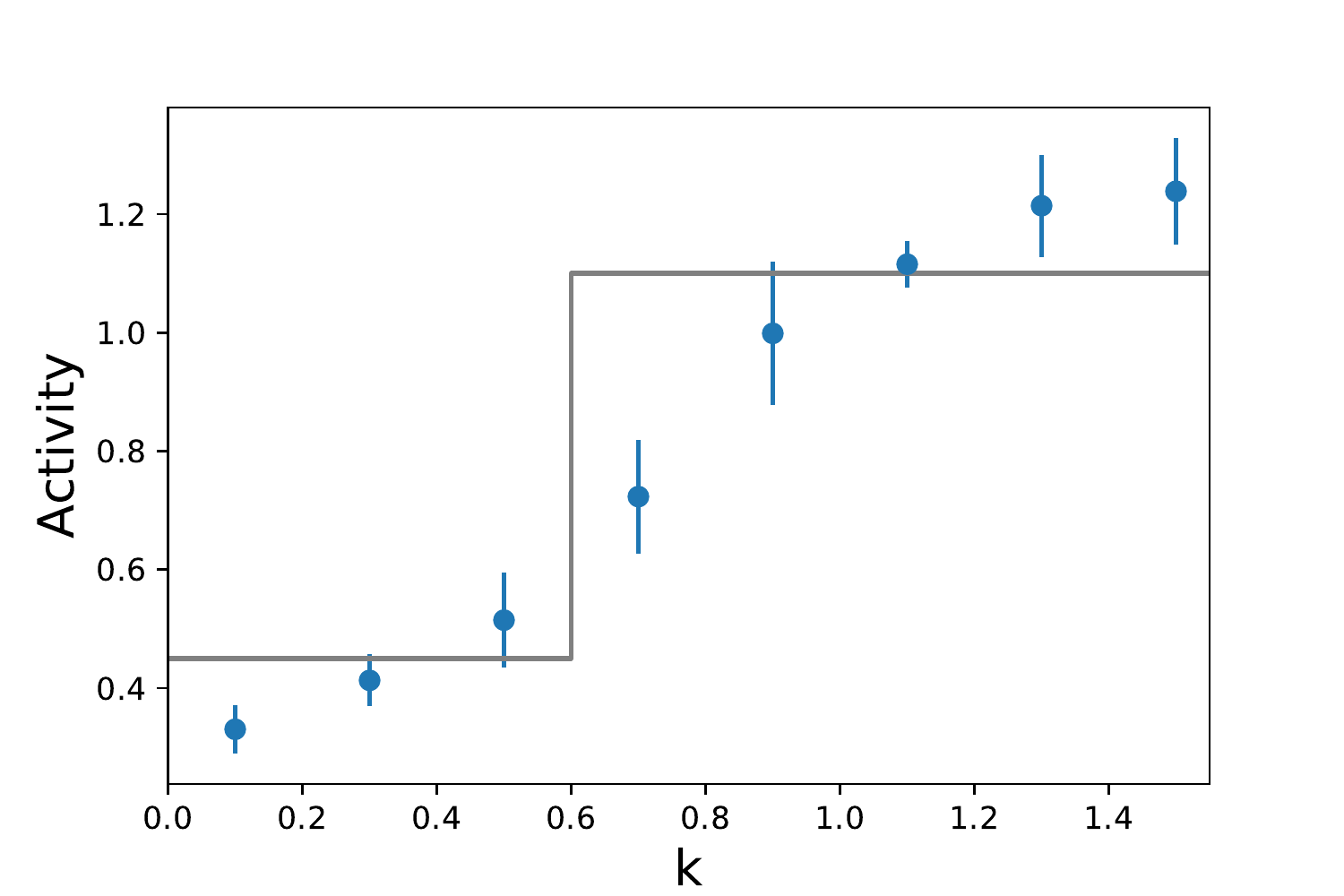}
\caption{An example of the AND gate (Fig.~\ref{fig:andDemo}). The activity of output
neuron is nonlinear as function of the connection strength $k$.
It resembles a smoothed version of step function expected for
a discrete case.}
\label{fig:andDemoPlot}
\end{figure}
We are interested in two different modes of the system:
$\varepsilon_1 = \varepsilon_2 = \hat{\varepsilon} = 8000$ (On AND
On should result in On output) and $\varepsilon_1 = 0,
\varepsilon_2 = \hat{\varepsilon}$ (Off AND On should result in
Off output). We choose $\varepsilon_1 = \hat{\varepsilon},
\varepsilon_2 = k\hat{\varepsilon}$. The plot of the output neuron's activity as function of $k$ is shown in
Fig.~\ref{fig:andDemoPlot}. One may recognize a step function in the
output pattern.

\subsubsection{Logical NOT}

Up to this point we caused spikes in the neurons, but to
implement arbitrary logic a suppression is required. For
this purpose we introduce a logical NOT element. To this end one adds a certain type of connection called an \textit{inhibiting} one. Two neurons connected by such a connection should not spike simultaneously. The interaction part of the Lagrangian that implements the proposed behavior can be written as follows:
\begin{equation}
\mathcal{L}_{int} = \varepsilon_{inh}\left(\varphi_{1}^2 - 1\right)^4\left(\varphi_{2}^2 - 1\right)^4.
\end{equation}
%\begin{figure}
%\includegraphics[width=\linewidth]{inhibition.pdf} % first figure itself
%\caption{Inhibiting potential that is used to prevent simultaneous spike of %$\varphi_1$ and $\varphi_2$.}
%\label{fig:notPotential}
%\end{figure}
\begin{figure}
\includegraphics[width=0.7\linewidth]{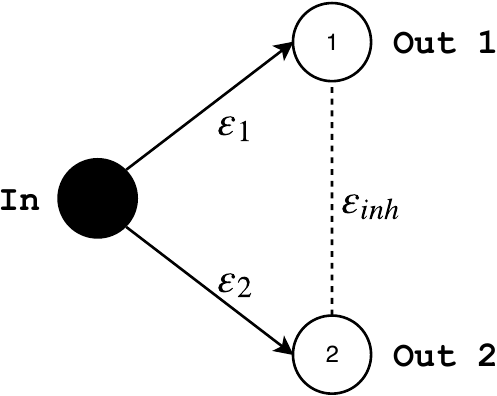} % second figure itself
\caption{A scheme implementing logical NOT. Due to $\varepsilon_{inh}$ only
one of output neurons can be active at a moment.}
\label{fig:inhibitingDemoScheme}
\end{figure}

\begin{figure}
\includegraphics[width=\linewidth]{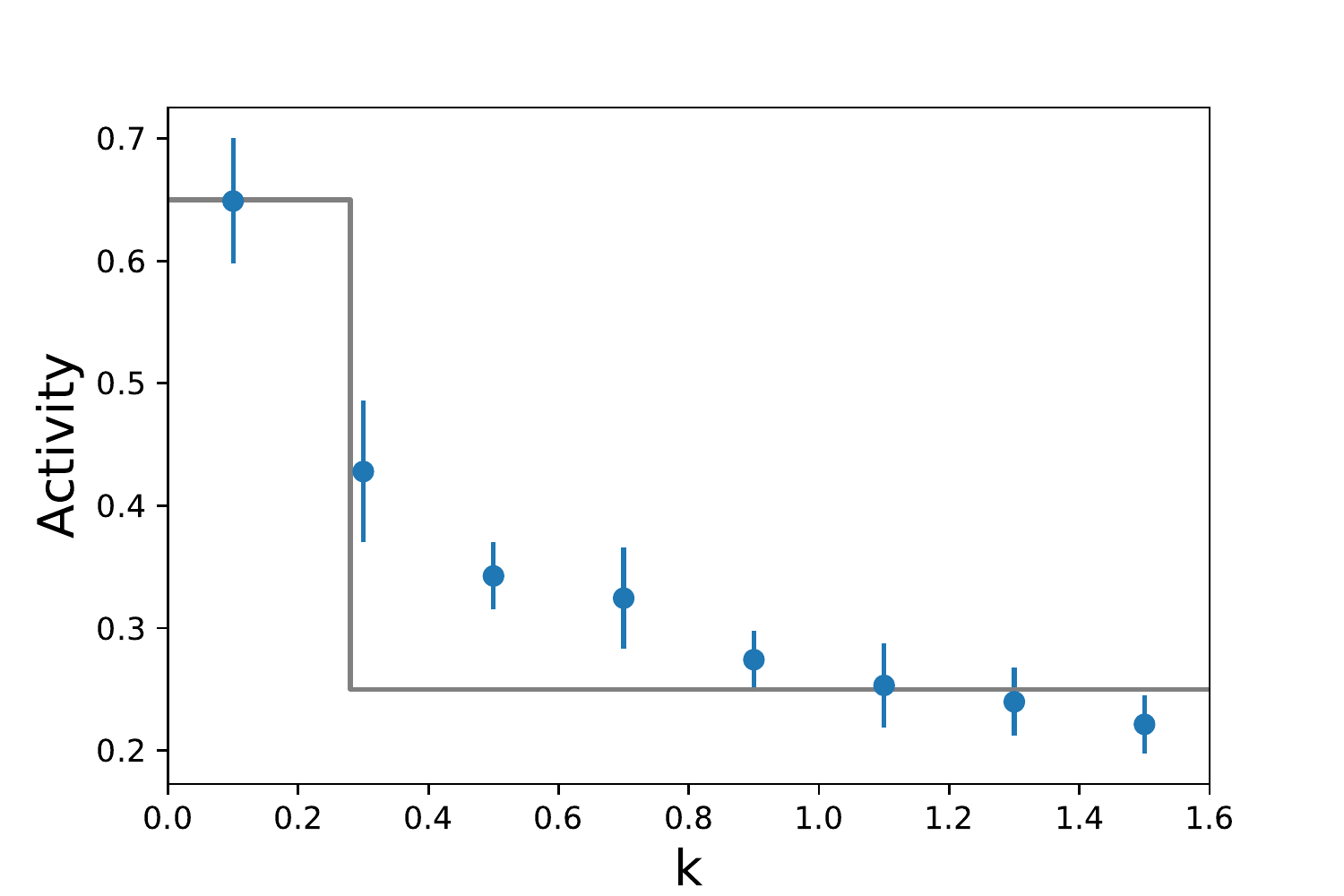}
\caption{The NOT gate (Fig.~\ref{fig:inhibitingDemoScheme}). The activity of output neuron 2 is shown. A neuron connected weaker to the input is inhibited by another neuron as inhibiting connection becomes stronger.}
\label{fig:logicalNOTexc}
\end{figure}
%The plot of the corresponding potential is shown in
%Fig.~\ref{fig:notPotential}. 
The interaction under discussion has an effect only if both neurons experience a spike.\par To demonstrate the operation of such a connection let us consider the scheme depicted in Fig.~\ref{fig:inhibitingDemoScheme}. A dashed line shows
the inhibiting connection. We set $\varepsilon_1 > \varepsilon_2$.
The activity of neuron 2 as function of $\varepsilon_{inh}$ is
presented in Fig.~\ref{fig:logicalNOTexc}, where
$\varepsilon_{inh} = 50000 k$ was set. As one can see, when
$\varepsilon_{inh} \approx 0$ neuron 2 is active however neuron 2 becomes inhibited by the neuron 1 as $\varepsilon_{inh}$ grows.

\subsubsection{Logical OR}

\begin{figure}
\includegraphics[width=0.7\linewidth]{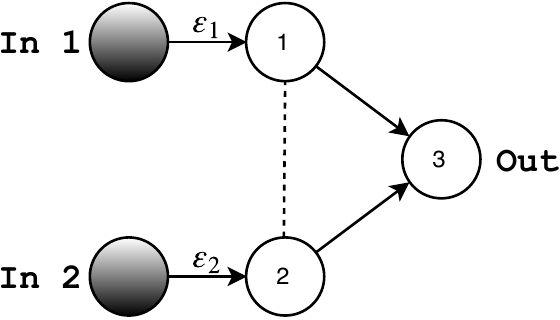} % first figure itself
\caption{An implementation of the logical OR gate.}
\label{fig:orScheme}
\end{figure}

Let us consider a logical OR element. The following behaviour is desired: the output neuron connected to a set of other neurons is activated when at least one of the input neurons becomes active. At a first glance it may seem that it is enough to simply
connect the neurons with an ordinary exciting connection with
a sufficient $\varepsilon_{exc}$. Unfortunately, this can not be performed: as noted above, an appropriate value of $\varepsilon_{exc}$ is restricted from above and this requirement would be violated if all the excited neurons  experienced a spike at the same time (effective $\varepsilon_{exc}$ is a sum of all the $\varepsilon_{exc}$'s of active neurons). A correct construction is depicted in
Fig.~\ref{fig:orScheme}. It turns out that intermediate neurons should be used to inhibit each other in such a way that only one of the elements is active at the same time. And it is these neurons that can be connected to the output neuron. Since the intermediate neurons become active one by one, the output neuron will never be overwhelmed.\par
%\begin{figure}
%\includegraphics[width=0.7\linewidth]{xor_scheme.pdf} % second figure itself
%\caption{The logical XOR implementation scheme. $\varepsilon_1 = 3000, %\varepsilon_2 = 10000.$}
%\label{fig:xorScheme}
%\end{figure}

\begin{figure}
\includegraphics[width=\linewidth]{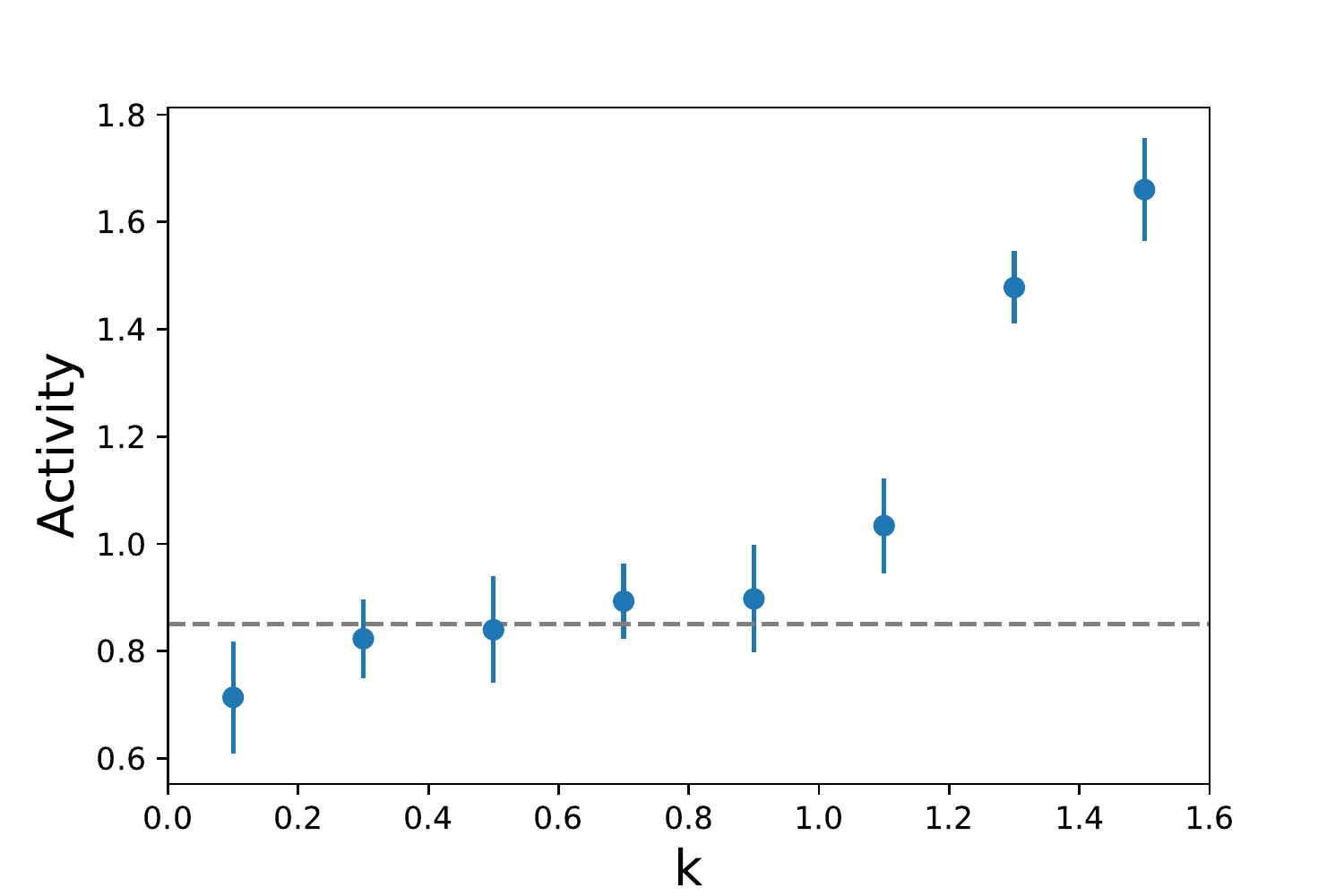}
\caption{The OR case (Fig.~\ref{fig:orScheme}). The activity of
output neuron 3. The output neuron is affected only by the neuron those input is the strongest. In the case of $k=1$ both the input neurons have an equal impact.}
\label{fig:logicalOR}
\end{figure}
Again let us consider two different modes of operation: $\varepsilon_1 =
\varepsilon_2 = \hat{\varepsilon} = 8000$ (On OR On should result
in On) and $\varepsilon_1 =0, \varepsilon_2 = \hat{\varepsilon}$
(Off OR On should also result in On). We choose $\varepsilon_1 =
\hat{\varepsilon}, \varepsilon_2 = k \hat{\varepsilon}$ and plot
the dependency of the output neuron's activity on $k$
(Fig.~\ref{fig:logicalOR}). Note that one may treat $OR(a, b)$ as
$max(a, b)$: $\forall a, b \in \{0, 1\}: max(a, b) = OR(a, b)$.
Our expectations are confirmed by the plot. The activity of
the output neuron does not change ($activity(max(\hat{\varepsilon},
k\hat{\varepsilon})) \overset{k < 1}{=}
activity(\hat{\varepsilon})$) when $k < 1$ and the activity of output neuron also increases as $k$ becomes bigger than 1.

%\subsubsection{An example of construction of logical XOR}

% We already have elements from which we can build arbitrary logic,
%but before we move on to more complex schemes, we will make sure
%that those elements can operate together. To do this, we construct
%from them a slightly more complex element - the exclusive or. The
%idea of its operation is quite expected: the output neuron should
%be active if and only if one of the input neurons is active. \par
%We achieve this by means of the scheme depicted in
%Fig.~\ref{fig:xorScheme}. Operation of the scheme is very simple:
%neuron 4 is active when at least one of the input neurons is
%active. But if both are active, neuron 1 will inhibit its
%activity. \par
%\begin{figure}
%\centering
%\includegraphics[width=\linewidth]{xor.pdf}
%\caption{The XOR case (Fig.~\ref{fig:xorScheme}).  Activity of
%output neuron resembles smoothed step function: it is high if one
%of the input neurons is effectively turned off by damping its
%connections with small factor of $k$ and low if both neurons are
%active.}
%\label{fig:logicalXOR}
%\end{figure}
%Let's see how output neuron (4-th) react to change in input's
%$\varepsilon$. Activity of output neuron as function of $k_1$ is
%shown in Fig.~\ref{fig:logicalXOR}. If first neuron is off ($k_1 =
%0$) then activation of second neuron leads to increase of activity
%of output neuron. But if first input neuron is active then
%activation of second neuron leads to decrease of activity of
%output neuron.

\section{Applications of the model}

\subsection{Convolutional model for vertical line detection}

Let us consider a simple example of a complex system -- an entire neural network. We pay an ultimate attention to the convolutional neural network \cite{lecun} popularized by Krizhevsky et al. \cite{KrizhevskyConv}.\par The basic operation principle of such a network is the following: each pixel of an input image can be represented by a neuron with a varying activity depending on the input color. A convolution operation with a certain
predetermined kernel is applied to the input image. In our case
the kernel is a matrix of size $3\times3$ with elements that can
represent $\varepsilon_{exc}$ or $\varepsilon_{inh}$. We apply
this matrix element-wise to the input image in every possible
position. Depending on the position where the kernel is applied the neurons of the input layer will be connected to some neuron of the second layer.\par As an example, one simple application of a convolution model -- detection of a vertical line -- is discussed below. Let us formalize the problem as follows: at the input we have a picture of size $4\times4$ pixels. The network is supposed to be able to detect a vertical line in this picture. If there is something not resembling a vertical line in the input image, the network should not react. Since the input layer has dimensions $4\times4$, the second one should be of size $2\times2$ and the third layer is represented by a single neuron. We connect all neurons of the second layer to the neuron of the third one by using $\varepsilon_{exc} = 4000$. The following kernels were used to connect the first and the second layers:

\begin{equation}
K_{exc} = \begin{pmatrix}
              0 & 1 & 0\\
              0 & 1 & 0\\
              0 & 1 & 0
             \end{pmatrix} \times 2000, 
\end{equation}

\begin{equation}
K_{inh} = \begin{pmatrix}
              1 & 0 & 1\\
              1 & 0 & 1\\
              1 & 0 & 1\\
            \end{pmatrix} \times 15000.
\end{equation}

\begin{figure}
\includegraphics[width=\linewidth]{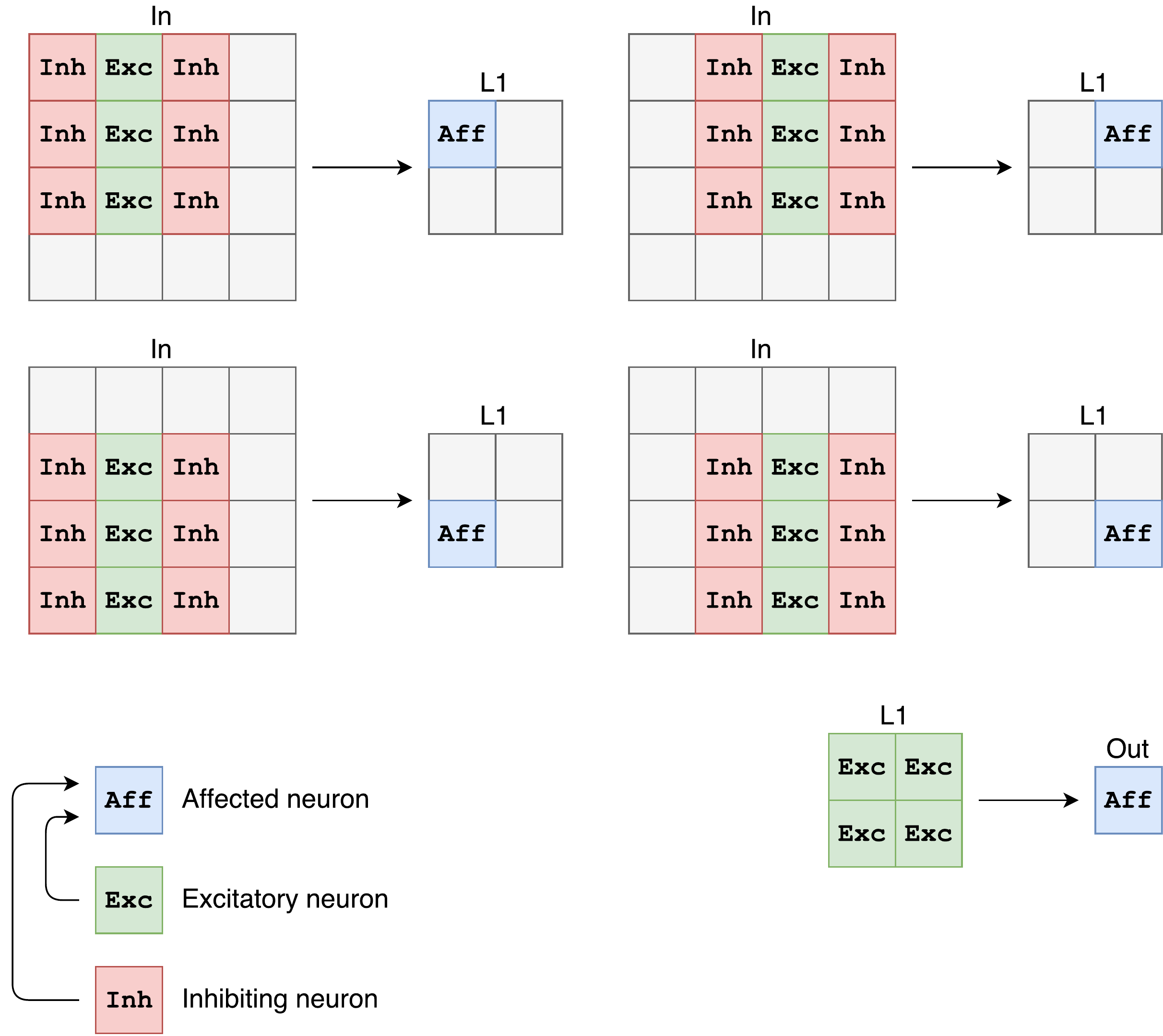}
\caption{Proposed architecture of a convolutional network.}
\label{fig:convScheme}
\end{figure}

\begin{figure}
\includegraphics[width=\linewidth]{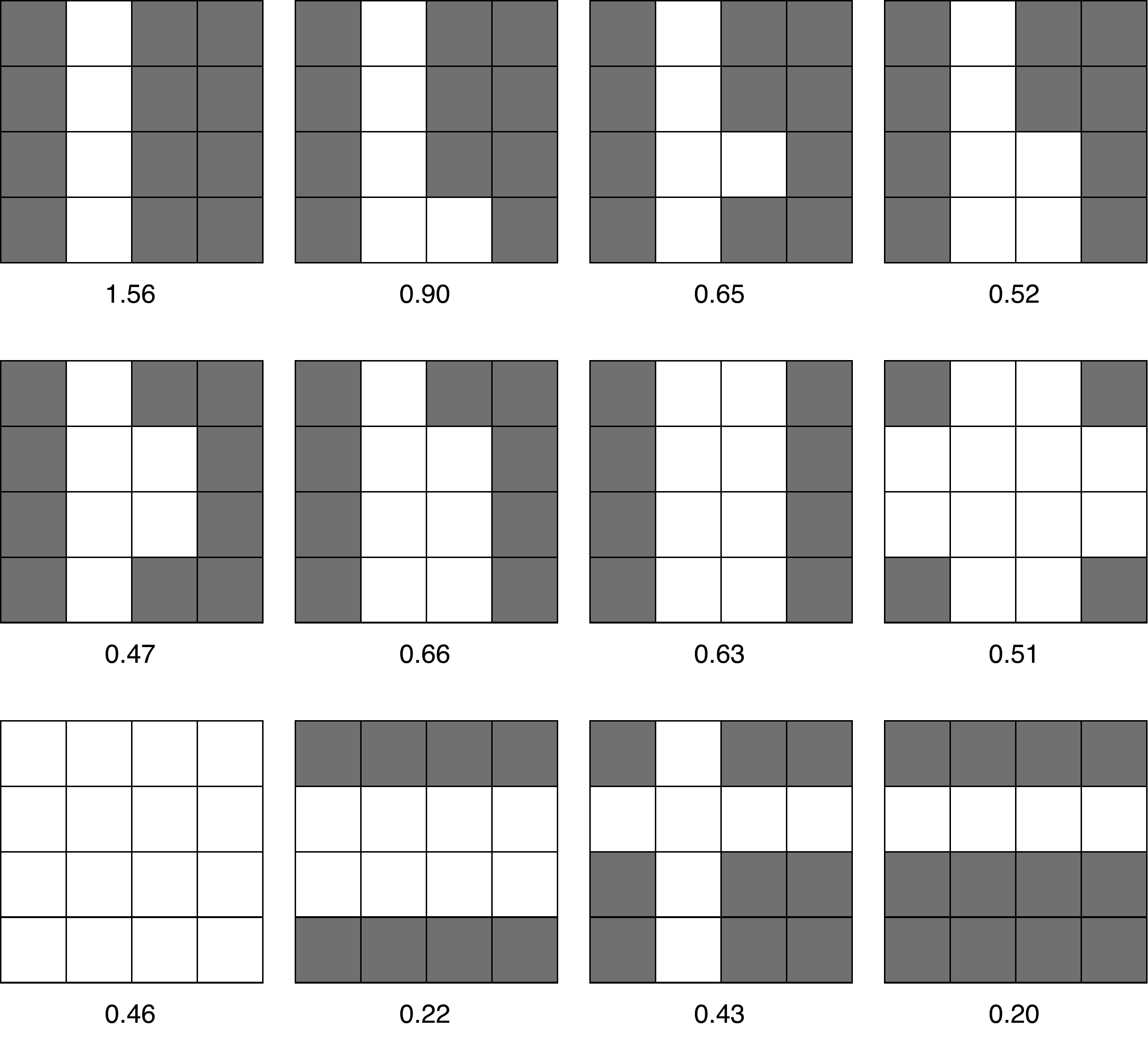}
\caption{Activity of output neuron is high in case if input resembles vertical line and low otherwise.}
\label{fig:lineDetection}
\end{figure}

In Fig.~\ref{fig:convScheme} the general scheme of the
network proposed is depicted, and in
Fig.~\ref{fig:lineDetection} different input images and corresponding network reactions are shown. As one can see, the network performs well and finds a vertical line safely.\par When dealing with real problems an input image will have much larger dimensions, so that the number of layers have to be increased to combine different kernels accordingly and ending up with increasingly complex images. A simulation of such a network can require long working times, but it will not have any fundamentally new parts. Thus, we can conclude that the model proposed can serve for implementing a convolutional neural networks.

\section{Conclusion}

%%A neural network model based on the behavior of interacting particles in a $W$ potential was proposed. Neuron connections of two types (excitation and inhibitory) were introduced. It was shown that the simplest logical elements can be implemented using such neurons. The simplest convolutional neural network for a vertical line recognition was constructed. It is demonstrated that an implementation of classical neural network on the base of the quantum system in the case of simple architectures is possible.\par
In the work we demonstrate that the system of interacting quantum particles in $W$ potential can serve as artificial stochastic neural network. Stochastic neurons in this approach are the particles in the potential and stimulated instantons play the role of spikes of activity.

We propose a kind of connection between neurons, which leads to the fact that a spike of activity on one neuron stimulates a spike on the associated neuron. Thus, neurons connected in a chain propagate activity along it. An essential property of the model is a nonlinear dependence of the ability to spread activity on the magnitude of the connection. So, activity is not transmitted until the critical value is reached, and then it increases sharply. These facts allow us to propose a scheme for the implementation of logical elements based on a network of neurons. Our work also describes an inhibitory relationship that is used for logic OR and NOT gates. A convolutional neural circuit can also be built, which is demonstrated on the example of a network for recognizing a vertical line.

An interesting question beyond the scope of the presented research is a practical implementation of the mentioned neural network. Note that a quantum system of particles in the $W$ potential can be achieved by using a double quantum well. This has also been proposed for building quantum computers. We believe that stochastic neurons can be similarly implemented on the basis of such quantum systems. In addition, the issue of using the proposed stochastic neural network deserves special attention. We assume that the advanced connection options of the proposed logic elements should be applied for advanced tasks such as number recognition. The tools presented in the work give hope for a successful solution to this problem.

Besides, let us note an important feature of the model proposed. Fig. \ref{fig:LINE_3} shows that the transport of the activity along the chain of neurons resembles a phase transition. If the coupling constant is small, no activity is transmitted to the output neuron at all. On the contrary, if the value of the coupling constant exceeds a certain critical value, a sharp increase in activity on the output neuron is revealed. Thus, the proposed model is an example of a neural network with a phase transition. It is well known that such neural networks are extremely interesting due to an implementation of edge-of-chaos learning \cite{edge-of-chaos-ler1}, and of creating a Boltzmann machine capable of self-organization and adaptive self-learning \cite{bp_bm}. Such Boltzmann machines are based on the principle of self-organized criticality \cite{bak1}, \cite{bak2}. This principle is spread widely in natural science \cite{crit_RMF}, especially in biology \cite{edge-of-chaos_brain}. Also, it is actively discussed in the context of the implementation of neural networks \cite{edge-of-chaos-ler2}, \cite{edge-of-chaos-ler3}. The model proposed makes it possible to investigate the phenomenon of self-organized criticality both numerically and analytically in a framework of a simple quantum mechanical problem. Such studies will be the subject of our further research.

\section{Acknowledgements}

The authors would like to thank the anonymous reviewer for careful reading of the manuscript. The work was supported  by  grant  from  the  Russian  Science  Foundation  (project  number  21-12-00237).

\end{document}